\documentclass[dvips]{acta}
\usepackage{supertabular,lscape,epsfig}
\usepackage{amssymb}
\usepackage{amsmath}
\usepackage[T1]{fontenc}

\SetPages{0}{0}

\SetVol{76}{2026}

\usepackage{lmodern}

\newcommand{\DI}{\mathrm{DI}}
\newcommand{\QC}{\mathrm{QC}}
\newcommand{\RC}{\mathrm{RC}}
\newcommand{\Tr}{\mathrm{Tr}}

\begin{document}

\begin{Titlepage}

\Title{On the numerical reliability of nonadiabatic pulsation solutions in AGB envelopes}

\Author{Zalewski, J.,}
{Independent researcher \\
e-mail: jan.zalewski.a2@gmail.com}

\Received{May 28, 2026}
\end{Titlepage}

\Abstract{
We re-examine integration methods of linear nonadiabatic radial pulsations in the envelopes of AGB and post-AGB stars using the
subspace shooting formulation. In the
article the accuracy of the direct integration, the Riccati based method and
a continuous renormalization nonlinear integration method are examined. The last method is not widely used
in stellar pulsation studies and is therefore of interest in the present comparison.

A tracking transformation of the two independent solution vectors is introduced. It helps to improve the accuracy and stability of the
solution and can be readily implemented for any of these integration methods.

Minimum singular value maps are used to examine the structure of the pulsation spectrum over the complex-frequency plane and also to assess the fidelity with which the integration methods transport the solution subspace. These maps provide a practical alternative to the customarily used maps based on the determinant.

We use these maps together with the eigenfrequency spectra to establish mesh conditions required for the reliable transport of the solution subspace, especially in the nonlinear methods.

To measure the quality of the subspace transport a spillover measure is introduced. Using this measure it is also possible to verify
the placement of the inner boundary, which in the case of radial modes, cannot be guided by propagation criteria based on the Lamb frequency.

Visual inspection of the eigenfrequency spectra, the singular value maps, or the eigenmode
amplitudes and behavior in the deep envelope does not necessarily imply preservation of the transported-solution subspace.
The spillover measure provides a much clearer indicator.

This approach provides explicit quantitative diagnostics for assessing the reliability of strongly nonadiabatic
pulsation calculations in AGB/post-AGB stellar envelopes.}

{stars: AGB and post-AGB, stars: oscillations, stars: interiors, methods: numerical}

\section{Introduction}

We examine numerical reliability, convergence behavior, and failure modes of several subspace integration formulations for radial, highly nonadiabatic AGB envelope pulsations. Strongly nonadiabatic envelope pulsations naturally lead to stiff systems of differential equations with exponentially separated local solution branches.

Methods for solving pulsation equations in the envelopes of supergiant stars have been studied by Dziembowski (1977), Osaki (1977) using direct integration methods and Saio and Cox (1980) using relaxation methods. Glatzel and Gautschy (1992) introduced nonlinear integration method based on the Riccati transformation as a means of neutralizing the adverse numerical effects of the disparity of local eigenvalues of the pulsation matrix. Later studies used these methods, for example Saio et al. (2006) obtained results using relaxation methods and Riccati ones. Alternative integration methods intended to improve numerical stability and accuracy of stellar pulsation integrations and relying on Magnus expansion were introduced by Townsend and Teitler (2013). 

The present study is restricted to surface-to-interior integrations in extended AGB/post-AGB envelopes. Questions related to integration direction, midpoint matching, and center-to-surface formulations are important but are left for future investigation in order to keep the present work focused on the numerical behavior of envelope integrations. The interplay between integration direction, midpoint selection, and subspace contamination may be of considerable importance in deep integrations and deserves separate investigation.

In the present paper we will focus on the integration of linear, radial pulsation nonadiabatic equations in the envelopes of AGB and post-AGB stars. We reformulate the approach in terms of subspace propagation, as this permits to introduce useful methods and measures.

In addition to the direct integration method and the Riccati method we implemented also the continuous renormalization nonlinear method and a tracking transformation of the integrated solutions improving their conditioning.

While the search for nonadiabatic pulsation frequencies is made difficult for strongly nonadiabatic modes, particularly for the strange modes, numerous methods were invented to aid in finding pulsation frequencies. For example Aikawa (1993) used a homotopy/continuation transformation enabling the determination of nonadiabatic modes from adiabatic ones, and Goldstein and Townsend (2020) used contour methods to localize pulsation frequencies based on the winding numbers for the complex pulsation determinant. We have implemented a map based on minimum singular value of the determinant matrix to obtain an overview of eigenfrequencies in the complex-$\sigma$ plane, and we use a grid of frequencies for direct frequency searches. Afterwards we compare the two.

Since one of the fundamental problems of envelope pulsations for radial modes is the lack of a well defined inner boundary we have implemented a criterion enabling placement of the inner boundary in the region where the solutions become asymptotically decaying. We also introduce geometric diagnostics of numerical reliability of pulsation integrations and use is to verify the placement of the inner boundary.

\section{Pulsation equations and boundary conditions}
Most of the notation that will be used in this paper was already introduced in Zalewski (2026). So here we will just summarize the basic elements that are of relevance for the present paper.

\subsection{Pulsation equations}
We use linear radial nonadiabatic pulsation equations in the notation of Dziembowski (1977) with Cowling approximation, which is applicable to extended envelopes of AGB/post-AGB stars. The dependent variables are: displacement \(d\), pressure \(p\), entropy \(s\) and luminosity \(f\) perturbations that form the solution vector \(y\). The pulsation equation 
\begin{equation}
	y'=M\ y
\label{eq:basicEq}	
\end{equation}
together with boundary conditions at the surface and at the bottom of the pulsation region enable to determine the (complex) frequency of pulsation and the eigenmode \(y(\sigma,x)\). The independent variable is \(x=\ln(r/R_{\odot})\). In what follows we will use non-dimensional frequency \(\sigma=\omega/\sqrt{4 \pi G \langle\rho\rangle}\). Since the time dependence of pulsation quantities is \(\exp(\omega\,t)\) thus \(\Im(\sigma)\) is the pulsation frequency, and the real part is the excitation rate.
 
\subsection{Boundary conditions and determinant construction}
These topics were discussed at length in Zalewski (2026). For the present article it is relevant that we will consider the outer and inner boundary conditions as determined by eigenvectors of the local pulsation matrix at the boundary to obtain initial values for the downward integration from the surface, as well as the left eigenvectors representing the unwanted (excluded by inner boundary conditions selector) subspace at the inner boundary and use it to derive the pulsational determinant from the matrix 
\begin{equation}
	B = (b_{i,j})_{i,j=1,2}
\label{eq:matrixB}
\end{equation}
with \(b_{i,j}=w^{\dagger}_i y_{b,j},\ i,j=1,2\) where \(w\) are the left eigenvectors of the pulsation matrix at the inner boundary spanning the unwanted subspace, and \(y_b\) are the two independent solutions' vectors at the inner boundary. 
This formulation enables us the control over boundary conditions and thus helps to obtain information about the solution and its properties. Since at the eigenfrequencies of Eq.~(\ref{eq:basicEq}) the \(\det(B)=0\) this property is used to iterate the pulsation frequencies starting from trial values.

Unless otherwise noted we will use the following boundary conditions selector \((2,4)\text{--}(1,3)\).

\subsection{Eigenfrequency maps using minimum singular value of \(B\)}
While it is customary to plot maps of \(|\det(B)|\) over complex-\(\sigma\) plane we adopt a different approach viz. we use the minimum singular value \(s_{min}\)  of matrix \(B\). The \(s_{min}(B)\) has zeros at the same frequencies that the \(\det(B)\) does. Additionally the singular value is real and thus more suitable for plotting of maps over complex-sigma plane. We introduce
\begin{equation}
	\delta \equiv -\log_{10}\ s_{min}(B)
\end{equation}
We use negative sign because then the zeros of \(B\) are represented as peaks in the map.

The \(\delta(\sigma)\) maps provide information not only about the pulsation spectrum, but also about the numerical conditioning of the boundary-value problem over the complex frequency plane, as will be discussed later on in the paper. 

\section{Integration methods}
The pulsation equation Eq.~(\ref{eq:basicEq}) is integrated using Runge--Kutta method RK4, wherein the coefficients of the pulsation matrix \(M\) are linearly interpolated between layers at the required intermediate points. The solution is propagated from surface to the inner boundary, layer by layer, using one of the integration methods discussed in the following subsections. 

All the methods considered here are one-sided shooting methods, which means they are applicable to the extended envelopes of AGB stars but their reach into the deep interior of the envelope is limited by the well-known numeric cancellation effects of the solution \(y\) when it is obtained from the linear superposition of the two linearly independent solutions \(y_{i=1,2}\) that are integrated from the surface. In the deep lower regions of the envelope the pulsation matrix \(M\) becomes stiff, which leads to exponentially increasing components that dominate the solution and lead to severe cancellation of \(y=c_1 y_1 + c_2 y_2\) (see for example Towsend and Teitler (2013)).

As discussed in Zalewski (2026) the integrations are carried inward until the condition \(\tau_{th}/\tau_{dyn} \sim |\omega|\tau_{th} = \texttt{RC1}\) is satisfied. We typically use \(\texttt{RC1}=5\), as this allows the solutions to reach asymptotic regime suitable for application of inner boundary conditions. To examine the effects of pushing the solution into regions where the parasite components start to dominate it higher values of \(\texttt{RC1}\) will be also used. 

The three integration methods, discussed in following subsections, will be used to compare numerical stability, computed frequency spectra and the behavior of the propagate pulsation eigenmodes.

\subsection{Direct integration - $\DI$}
This method employs direct integration of \(y\) from Eq.~(\ref{eq:basicEq}) using RK4. This method suffers from exponential instability if the integrations are continued too deep into the envelope, but is simple to implement. This method integrates the independent solutions of Eq.~(\ref{eq:basicEq}) thus it is easy to observe the behavior of these two solutions.

\subsection{Riccati integration - $\RC$}
This method was introduced by Glatzel and Gautschy (1992). It relies on a Riccati transformation of Eq.~(\ref{eq:basicEq}), leading to a nonlinear system describing the propagation of solution subspace \(u\) through the associated Riccati matrix \(R\), which represents ratios between selected solution components.

In our formulation, the resulting nonlinear equations for \(u'\) and \(R'\) are integrated layer by layer. At each layer, the determinants corresponding to the possible representations for \(u\) in terms of the dependent variables \(\{d,p,s,f\}\) and evaluated, and the pair maximizing the determinant is selected. This procedure thus maximizes the local mapping area and improves numerical stability. The selected representation is retained as long as no better-conditioned mapping is available. The remapping may occur at each layer.

The computed \(u_i\), \(R_i\) and mappings \(m_i,\ i=1,N\), where \(N\) is the number of layers are stored and used in subsequent reconstruction of the eigenmode components \(y_{i,j} \ i=1,N, j=1,2\) and of the mode.

\subsection{Continuously renormalized integration - $\QC$}
As a third integrator method we consider a nonlinear method relying on a continuous renormalization of the base, with refactoring and normalization to stabilize the solution variables. 

The method, described here, is closely related to continuous orthogonalization schemes (Drury (1980) and Davey (1983)) used in stiff linear boundary--value problems and Lyapunov exponent computations. The present formulation is closely related to the polar-coordinate and continuous orthogonalization approach described by Humphreys and Zumbrun (2005). 

Let the 
\begin{equation}
Y(x) = [y_1(x), y_2(x)]
\label{eq:bigY}
\end{equation} 
represent the solution matrix composed of the two independent solutions of Eq.~(\ref{eq:basicEq})  propagated from the outer boundary, then in the QC formulation the matrix \(Y\) is factorized as:
\begin{equation}
	Y = QC
\end{equation}
where \(Q\) is a \(4 \times 2\) matrix whose columns form an orthonormal basis of the propagated two-dimensional solution subspace,
\begin{equation}
	Q^H\,Q=I
\end{equation}
and \(C\) is a \(2 \times 2\) complex coefficient matrix carrying the scaling and mixing of the solutions within that subspace.

Substituting \(Y=QC\) into Eq.~(\ref{eq:basicEq}) gives
\begin{equation}
	Q'\,C+Q\,C'=M\,Q\,C.
\end{equation}

Defining 
\begin{equation}
	A=Q^H\,M\,Q
\end{equation}
and requiring preservation of orthonormality of the columns of \(Q\), the equations may be written as (see Humphreys and Zumbrun (2005), Eqs.~(8) and (10))
\begin{align}
	Q'&= M\,Q-Q\,A \\
	C'&= A\,C.
\label{eq:QCeqs}	
\end{align}
The matrix \(A\) represents the projection of the full operator \(M\) onto the propagated two-dimensional solution subspace. 

The equations for \(Q\) and \(C\) are integrated using RK4. After each step the columns of \(Q\) are re-orthonormalized using modified Gram--Schmidt orthogonalization. The corresponding upper-triangular transformation matrix \(R\) is absorbed into \(C\):
\begin{align}
	Q_{old} &= Q_{new} R, \\
	C &\leftarrow R\,C
\end{align}
thus preserving the fundamental solution matrix \(Y=Q\,C\).

In addition \(C\) is periodically re-factorized using unitary transformation
\begin{equation}
	C = U\,T
\end{equation}
where \(U\) is unitary and \(T\) - upper triangular. The transformation is transferred back to \(Q\)
\begin{equation}
	Q \leftarrow Q\,U,\  C \leftarrow T
\end{equation}
which preserves the propagated solution matrix while improving numerical conditioning.

To prevent numerical over/underflow associated with exponential growth of the coefficients of the matrix \(C\) an additional normalization step is used during integration. The matrix \(C\) is rescaled using its Frobenius norm \(D=\|C\|_F\), as \(C\leftarrow C/D\), and the scalar thus removed is accumulated in a logarithmic form as \(\ln S \leftarrow \ln S+\ln D\). Thus the final form of the solution matrix may be written as:
\begin{equation}
	Y=Q\,C\,e^{\ln S}.
\end{equation}

The QC formulation is intended to prevent rapid loss of linear independence deep in the envelope and thus offer better performance over direct integration. 

The performance of the integration methods presented will be discussed in subsequent sections following the next one which introduces a modification of the integration methods aimed to improve stability of the solution, particularly in regions where linear superposition of \(Y\) into \(y\) may lead to significant cancellation.

\section{Tracking transformation}
All the three integration methods introduced in the preceding section operate within the same framework in which two independent solutions are determined at the surface, using outer boundary conditions, and they are (or alternatively the subspace they span is) integrated to the inner boundary, where two other boundary conditions are used to define the wanted subspace. This enables the computation of the determinant and solution mixing coefficients \(c\) from \(B\,c=0\). Setting \(c_1=1\) the solution \(y(x)=y_1(x)+c_2\,y_2(x)\) is obtained. At the inner boundary it is typically found that
\[ \max(|y_1|,|c_2y_2|)/|y|> 10^5,  \] while on the average the value of this ratio is about $10^7$, thus the reconstruction results in large accuracy loss.

Motivated by the large cancellation between the propagated basis \(y_{i=1,2}\) we introduce a tracking basis as defined by \([y_1,y_2]\rightarrow [y_1, y^{\Tr}_2]\) where \(y^{\Tr}_2=y_1+c_2\,y_2\). Thus we integrate the subspace defined by \([y_1,y^{\Tr}_2]\) instead of the usually propagated subspace of \([y_1,y_2]\). The rationale for this method may be given as follows.

At the surface the outer boundary condition provides two independent vectors \(v_1,v_2\), which in our case are the eigenvectors of the local pulsation matrix determined by the adopted selector. Instead of setting \(y_1,y_2\) to these values, as is customary, we set \(y_1=v_1\) and \(y^{\Tr}_2=v_1+c_2\,v_2\). Since, at the surface, both \(|v_i|\sim O(1),\ i=1,2\) and the solution \(|y|_{surface}\sim 1\) then there does not occur substantial cancellation. Thus both \(y_1\) and \(y^{\Tr}_2\) are determined by the outer boundary condition with comparable accuracy.

Next we integrate \(Y^{\Tr}=[y_1,y^{\Tr}_2]\) to the inner boundary as described in the preceding sections. The difference from standard approach is such that once the pulsation frequency is determined, the eigenmode is not reconstructed from the linear combination of the basis vectors, but it is given explicitly by \(y=y^{\Tr}_2\). Thus, while accuracy may be lost due to the adopted integration method, we do not loose accuracy by reconstructing the solution from basis vectors.

Thus in this approach it is not only the frequency that is iterated but at the same time the eigenmode is obtained in the same process.

The use of the tracking method imposes following alterations to the standard operational procedure of obtaining solutions of pulsation equations. 

Since the value of \(c_2\) is needed at the start of integrations, together with initial guess for \(\sigma\), it needs to be assumed (we assume initial value of \(c_2=1\)). After the solution is propagated to the inner boundary a new value for \(c^{(n+1)}_2\) is obtained alongside an improved estimate of \(\sigma^{(n+1)}\) and the process is repeated until convergence. The iterations are stopped when the accuracy constraints on the changes in one step become satisfied: \(\Delta |\sigma|<\epsilon_{\sigma}\) and \(\Delta |c_2|<\epsilon_{c_2}\), with \(\epsilon\sim 10^{-10}\).

There are two more aspects of the computation that need to be addressed when using tracking method - viz. the scaling of the determinant, and the iteration of \(c_2\). The tracked and original bases are related through
\begin{equation}
	Y^{\Tr}=Y\,Tr
\end{equation}
with the transformation matrix
\begin{equation}
	Tr = 
	\begin{pmatrix}
		1&1\\
		0&c_2^{(n)}
	\end{pmatrix}
\end{equation}
the determinant of which is \(\det(Tr)=c^{(n)}_2\).

Therefore
\begin{equation}
	\det(B^{\Tr})= \det(B)\,\det(Tr)=c^{(n)}_2\det{B}.
\end{equation}

Hence the tracked determinant contains a scaling induced by the basis transformation and not related to the physics of the mode itself. The adopted scaling \(\det^{(n)}(B)= \det(B^{\Tr})/c^{(n)}_2\) eliminates that dependency. 

In order to obtain the value of \(c_2^{(n+1)}\) for the next iteration, it should be noted that the matrix \(B\) is computed from the tracked basis. At the inner boundary the condition \(B^{\Tr}\, c=0\) leads to 
\begin{equation}
\begin{pmatrix}
	w^{\dagger}_1 y_1 & w^{\dagger}_1 y^{\Tr}_2\\
	w^{\dagger}_2 y_1 & w^{\dagger}_2 y^{\Tr}_2
\end{pmatrix}\,
\begin{pmatrix}
	c_1 \\
	c_2
\end{pmatrix}=0.
\end{equation}

Thus from the first row it may be obtained (assuming \(w_1\neq 0\)) that \(c_1\, y_1 + c_2\,y^{\Tr}_2=(c_1+c_2)\,y_1+c_2\,c^{(n)}_2\,y_2=0\). Dividing by \(c_1+c_2\) and assuming \(c_1=1\) it is obtained that the mixing coefficient \(c^{(n+1)}_2\) is given by
\begin{equation}
c_2^{(n+1)}
=
c_2^{(n)}
\frac{c_2}{1+c_2}.
\end{equation}

\subsection{Application of tracking to integrators}
While it is obvious that the tracking may be applied to the direct integration method, thus obtaining $\DI^{\Tr}$, it may also be applied to obtain $\QC^{\Tr}$ and $\RC^{\Tr}$ methods.

For the $\QC^{\Tr}$ and $\RC^{\Tr}$ the essential changes occur outside of the integrator, viz. in the outer boundary conditions formulation, and the determinant and \(c_2\) computation. In both cases the solution is obtained from the second vector \(y^{\Tr}_2\). The first one serves only in the derivation of matrix \(B\) for all three methods.

\section{Spectral structure and numerical convergence}
In this section we will present and discuss the behavior of frequency spectra for the following cases of meshes with logarithmic step size and meshes with variable step sizes.

\subsection{Constant step integrations}
The envelope model is computed using a logarithmic steps defined as: \(\Delta x=\alpha_c\, \ln(R_{surf}/R_{\odot})\), with a fixed \(\alpha_c\). We have computed three envelope models with parameters \(\log(T_{eff}=3.85)\), \(\log(L/L_{\odot})=4\), \(M/M_{\odot}=0.69\) and \(\alpha_c=\{\texttt{5E-4},\ \texttt{5E-5},\ \texttt{5E-6}\}\). With these parameters there were respectively \texttt{3.9k}, \texttt{39k}, \texttt{390k} layers in the model.

In order to visualize the numerical behavior of the integrators, we have computed the \(\delta\)-maps over complex frequency range: \(-3\leq\Re(\sigma)\leq 3\) and \(0\leq\Im(\sigma)\leq20\). This frequency range covers p-modes up to about 20\textsuperscript{th} overtone, and the wide excitation rate range covers strange modes. Thus the maps should contain p-modes and strange modes.

\begin{figure}[htb]
	\includegraphics{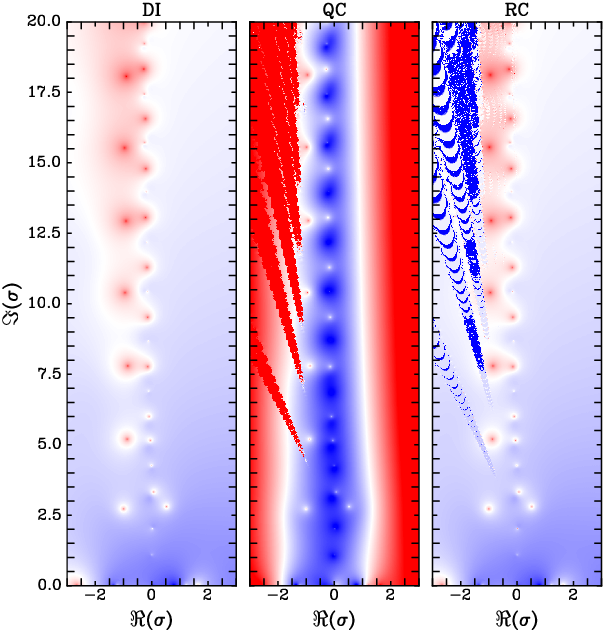}
	\FigCap{\(\delta\)-maps for the \(\alpha_c=\texttt{5E-4}\) model. Left pane - map of \(\delta(\sigma)\) for the $\DI$ integrator, middle panel: $\QC$, right panel: $\RC$. The nonlinear integrators $\QC$ and $\RC$ show ridge structures absent for $\DI$.}
	\label{fig:Fig1}	
\end{figure}

Such maps are presented on Fig.~\ref{fig:Fig1} where the \(\delta(\sigma)\) is shown for the $\DI$, $\QC$ and $\RC$ integrators. As may be seen from Fig.~\ref{fig:Fig1} there appear artifacts in the \(\delta\) maps for the nonlinear integrators $\QC$ and $\RC$. These artifacts emerge at approximately the same locations for these two methods: \(\Re(\sigma)<-1\) and \(\Im(\sigma)>5\) for this model. In case of the $\QC$ integrator the artifacts form valleys in the map of \(\delta(\sigma)\) while for the $\RC$ one they form ridges. On the map for $\RC$ one may also see wobbles in the \(\delta\) at high frequencies (\(\Im(\sigma)>13\)). In what follows these forms of the artifacts will be referred to as ridges. The ridges may reflect the distinct nonlinear character of the $\QC$ and $\RC$ formulations.

However the pulsational frequencies found for $\DI$ are also recovered by $\QC$ and $\RC$.

\begin{figure}[htb]
	\includegraphics{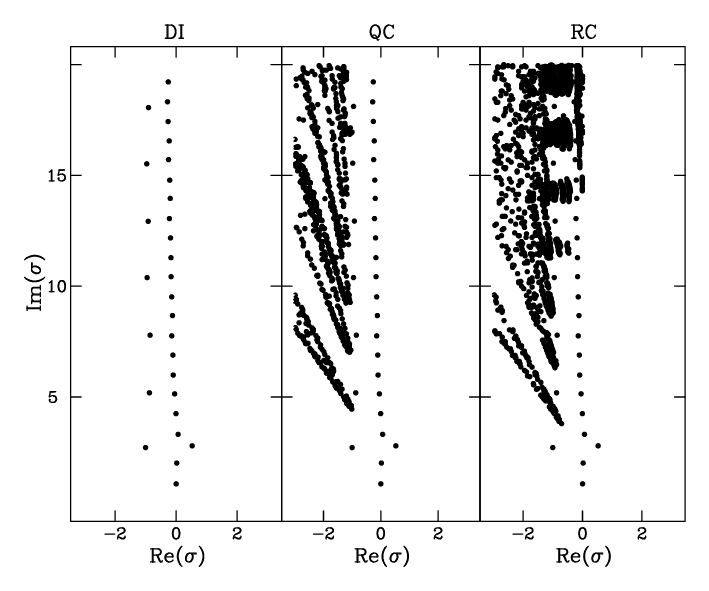}
	\FigCap{Frequency spectra for the \texttt{5E-4} model, and for DI, QC and RC integrators. The nonlinear integrators show artifact frequencies in the general area of the ridges seen in Fig.~\ref{fig:Fig1}. Strange modes are clearly seen for this model in the left pane for the $\DI$ integerator; they occur for $\Re(\sigma)<-0.5$ for the damped modes. A single excited strange mode ($S^+_1$) is seen near $\Im(\sigma)=2.9$.}
	\label{fig:Fig2}
\end{figure}

The effect these three integration methods have on pulsation frequencies may be seen in Fig.~\ref{fig:Fig2} where the results of pulsation frequency search are presented for the same model as in Fig.~\ref{fig:Fig1}. It may be seen that artifact frequencies are present for both nonlinear integrators, and they group in the area of the ridges. In case of Riccati based integrator the spurious frequencies surround the strange modes and extend towards positive \(\Re(\sigma)\). The ridges and wobbles in the \(\delta\) map for $\RC$ are thus the areas where artifact frequencies are found. It should be noted that our integration methods entail integration from surface to the inner boundary and this may contribute to the emergence of the artifacts, though, as it will be shown later, the direction of the integration alone is not sufficient to explain the artifacts.

It should be stressed that the modes corresponding to the spurious frequencies may appear regular, exhibiting exponential amplitude decay towards the interior and converged values for the kinetic ($EK$) and dissipation ($ED$) energy integrals, while others show clearly unacceptable behavior. Thus, in an automated frequency survey, the mode structure, amplitudes and convergence properties must be analyzed in order to ascertain whether a particular frequency corresponds to a physical eigenmode or to a numeric artifact. Some of the diagnostics that may be used for this purpose will be discussed in later sections.

In order to facilitate further discussion, we introduce a measure of the amount of artifact frequencies present in a spectrum as:
\[
C=\frac{N_{\rm spur}}{N_{\rm tot}}
\]
where \(N_{\rm tot}\) is the total number of frequencies found in the considered range using a given integration method, and 
\[
N_{\rm spur}=N_{\rm tot}-N_{\rm eig},
\] 
with \(N_{\rm eig}\) denoting the number of persistent eigenfrequencies identified by comparing the spectra for the three methods, and from convergence under mesh refinement. The \(0\leq C\leq 1\), with \(C=0\) meaning there are no artifact frequencies. For the case in Fig.~\ref{fig:Fig2} the values of \(C\) are: 0, 0.97 and 0.98 respectively.

The results regarding \(C\) for the two remaining \(\alpha_c\) values are given in Table~\ref{tab:contamination}. As it may be seen the artifact ridges disappear for the $\RC$ method only for the smallest fixed step size of \(5\times 10^{-6}\) with a mesh size of \texttt{390k}, while the $\DI$ integration method produces stable results for \(5\times 10^{-4}\) and with \texttt{3.9k} layers.

\MakeTable{lcllllc}{12.5cm}{Comparison of mesh types, number of layers and contamination factor \(C\)\label{tab:contamination}}
{
		\hline
		Method & Mesh             & \(\alpha_c\)&\(\alpha_v\) &\(\alpha_k\) & Layers & \(C\) \\
		\hline
		DI     & fixed            &\texttt{5E-4}&0            &0            & 3.9k   & 0.00  \\
		QC     & fixed            &\texttt{5E-4}&0            &0            & 3.9k   & 0.97  \\
		RC     & fixed            &\texttt{5E-4}&0            &0            & 3.9k   & 0.98  \\
		QC     & fixed            &\texttt{5E-5}&0            &0            & 39k    & 0.79  \\
		RC     & fixed            &\texttt{5E-5}&0            &0            & 39k    & 0.50  \\
		QC     & fixed            &\texttt{5E-6}&0            &0            & 390k   & 0.00  \\
		RC     & fixed            &\texttt{5E-6}&0            &0            & 390k   & 0.00  \\
		\hline
		QC     & mixed            &\texttt{5E-4}&\texttt{1E-2}&0            & 7.6k   & 0.51  \\
		RC     & mixed            &\texttt{5E-4}&\texttt{1E-2}&0            & 7.6k   & 0.92  \\
		QC     & mixed            &\texttt{5E-4}&\texttt{5E-3}&0            & 15k    & 0.00  \\
		RC     & mixed            &\texttt{5E-4}&\texttt{5E-3}&0            & 15k    & 0.38  \\
		QC     & mixed            &\texttt{5E-4}&\texttt{1E-3}&0            & 89k    & 0.00  \\
		RC     & mixed            &\texttt{5E-4}&\texttt{1E-3}&0            & 89k    & 0.00  \\
		\hline
		QC     & phase-controlled &\texttt{5E-4}&\texttt{5E-3}&\texttt{1.0} & 33k    & 0.00  \\
		RC     & phase-controlled &\texttt{5E-4}&\texttt{5E-3}&\texttt{1.0} & 33k    & 0.00  \\
		\hline
}

\subsection{Variable step integrations}
The examination of the artifact eigenmodes found for $\QC$ or $\RC$ nonlinear integrators shows that the under-sampling in the H/He ionization zones affects these methods substantially. Therefore we have implemented variable step integration mode whereby the tolerance is set on the amount of change from one layer to the next of the basic thermodynamic quantities (like pressure or temperature) supplemented by conditions on the coefficients \(\texttt{A}_{i=1,13}\) (Dziembowski, 1977) from which matrix \(M\) is computed. Thus for this kind of mesh the step size is controlled by \(|\Delta a/a|< \alpha_v\). The variable step size is halved if the condition on the previous step is not satisfied, and doubled if the change is less than half of the allowed tolerance \(\alpha_v\). The adaptive procedure could increase the step size beyond the fixed-step value, which may easily occcur in the outer layers, so the latter is imposed as an upper limit. The mixed step size control results in mesh that is dense in the ionization zones. This type of mesh is commonly used in pulsation studies, although conditions on the coefficients \(\texttt{A}\) are not usually enforced explicitly.

The results regarding \(C\) are given in Table~\ref{tab:contamination} and, as it may be seen, no artifact modes were found in the investigated frequency range for both nonlinear methods when \(\alpha_v=1\times 10^{-3}\). With this type of mesh the nonlinear integrators require \texttt{15k} layers for the $\QC$ integrator and about \texttt{89k} layers (for the model considered here) for $\RC$. This mesh performs thus substantially better than a fixed size one.

\subsection{Phase-controlled step integrations}
As a last option we considered a variable mesh with step-size controlled, not only by the preceding mechanisms but additionally, by the requirement that in one step the change of phase is controlled by the condition \(|k_{max}|\, \Delta x < \alpha_k\), where \(k_{max}\) is the largest magnitude eigenvalue, \(|k_{max}|=\max_i|k_i|,\ i=1,4\) of the locally computed matrix \(M\). In order to compute the coefficients of matrix \(M\) during envelope integration it is necessary to assume a certain maximum value of frequency \(\sigma\). In what follows we have assumed \(\Im({\sigma_{max}})=40\). The matrix \(M\) infinity norm is defined as \(||M||_{\infty} = \max_i \sum_j |M_{i,j}|,\ i,j=1,4\). With this norm the 
\begin{equation}
|k_{max}|\leq ||M||_{\infty}.
\label{eq:kMax}
\end{equation} 
The variable step size, as computed based on \(\alpha_v\), is modified by the condition resulting from \(\alpha_k\), thus ensuring the local oscillatory/exponential scale of the solution is sufficiently resolved. This kind of mesh leads to a decrease of step size deep in the envelope (as compared to the variable step size mesh) where the characteristic eigenvalues \(k\) become large because of the increasing thermal time scale relative to the dynamical one.

The results for this type of mesh are presented in Table~\ref{tab:contamination}. It was found that it is sufficient to require \(\alpha_k=1\), leading to \texttt{33k} layers to ensure the nonlinear integration methods do not produce artifact modes.

Therefore it follows that for the direct integration $\DI$ method it is sufficient (for the frequency range studied here) to use fixed mesh and about \texttt{4k} layers in the envelope to obtain stable eigenfrequencies and eigenmodes. For the continuous renormalization $\QC$ method a variable step size mesh is needed with about four times more layers. While for the Riccati based integration method $\RC$ a phase-controlled step size mesh is needed with about \texttt{33k} layers.

In the next section we will examine the behavior of modes so as to compare the methods and introduce quantitative measures of mode quality.

\section{Eigenfunction reliability and integration failure modes}
In this section we examine the reliability and integration failure modes of solutions of the radial pulsation equations in AGB/post-AGB envelopes. We compare the stability of eigenfrequencies obtained with different integration methods and mesh types, and analyze the behavior of eigenfunctions in the ionization zones and in the deep layers of the envelope. Finally, we introduce quantitative measures indicating impending loss of integration reliability.

\subsection{Stability of eigenfrequencies}
While the eigenfrequencies of AGB envelope pulsation modes are usually considered robust and resilient to the details of integration methods we intend to reexamine this using the three integration methods plus the tracking option and the different mesh types to verify if any of these factors impact the eigenfrequencies of p-modes or strange modes in the studied frequency range.

For this purpose we have selected the following four modes: the second overtone p-mode (2-ov), the strongly excited strange mode \(S^+_1\), the high-frequency mode (20-ov) and a damped, high-frequency strange mode (\(S^-_7\)) which is located close to the area where ridges appear in \(\delta\)-maps. The aim is to verify whether the eigenfrequencies, in the studied frequency range and for these mode types, are sensitive to these factors.

Table~\ref{tab:eigenfrequencies} lists the relative errors of eigenfrequency determination depending on the integration method and the mesh type used. We have used the average of the eigenfrequencies computed for the $\QC$ and $\RC$ methods on the phase-controlled mesh as a reference and measured the relative differences between an average of the given integration method and its tracked version and the reference value. Next we have averaged the relative errors for the four modes for real and imaginary parts separately:
\[
q_R = \frac{1}{N}\sum_{i=1}^N\left|\frac{\Re(\sigma_i)-\Re(\sigma_i^{\mathrm{ref}})}{\Re(\sigma_i^{\mathrm{ref}})}\right|,\ N=4
\]
and similarly for \(q_I\), with \(\sigma_i^{\mathrm{ref}}\) being the averages obtained for $\QC$ and $\RC$ integrations for the phase-controlled mesh.

In this way we have arrived at two numbers characterizing the quality of frequency determination for the given method and mesh type.

\MakeTable{lcccccc}{12.5cm}{Stability of the eigenfrequencies under different integration methods and mesh types\label{tab:eigenfrequencies}}
{
		\hline
		& \multicolumn{2}{c}{fixed}
		& \multicolumn{2}{c}{mixed}
		& \multicolumn{2}{c}{phase-controlled} \\
		\cline{2-3} \cline{4-5} \cline{6-7}
		
		method
		& \(q_R\) & \(q_I\)
		& \(q_R\) & \(q_I\)
		& \(q_R\) & \(q_I\) \\
		\hline
		
		DI
		& 1.7E-1 & 1.3E-2
		& 1.1E-3 & 4.7E-5
		& 3.6E-4 & 1.0E-5 \\
		
		QC
		&  & 
		& 8.0E-4 & 3.9E-5
		& 8.0E-7 & 6.4E-8 \\
		
		RC
		&  & 
		&  & 
		& 8.0E-7 & 6.4E-8 \\
		
		\hline
}

As may be seen from Table~\ref{tab:eigenfrequencies} the phase-controlled mesh provides most accurate results for both excitation rates and pulsation frequencies. This mesh has about twice as many layers as the mixed mesh and yet for $\QC$ the accuracy improvement from mixed to phase controlled mesh is three orders of magnitude. This indicates that phase controlled mesh offers a substantial advantage with regard to eigenfrequency determination. The equality of the QC and RC entries in the phase-controlled column results from defining the reference as their average.

The fixed mesh with fewest layers can be used with the $\DI$ method, however, the excitation rate is determined with substantially smaller precision, and even when the $\DI$ method is used with the phase-controlled mesh the accuracy is at best comparable to that of $\QC$ with mixed mesh. 

Thus it seems that the $\DI$ method with a sparse mesh may be used to obtain initial values for frequencies which can be used later on with $\QC$ or $\RC$ methods on a finer mesh to obtain more consistent values.

\subsection{Effects of tracking transformation}
While, over most of the envelope pulsation region the eigenfunctions such as \(|d|\) or \(|s|\) show practically no differences between the discussed integration methods, differences are expected to appear in the deep-envelope regions where pulsation equations admit both growing and decaying solution components.

It is for this reason that the inner boundary is determined using the parameter \(\texttt{RC1}=5\). This parameter is chosen so that the solutions of Eq.~(\ref{eq:basicEq}) enter the asymptotic regime but do not start to diverge. Such a choice is suitable for envelope pulsation modes studied here. With this choice of the parameter it is expected that the analyzed integration methods will produce practically identical eigenmodes, independent of whether tracking transformation is enabled. However when the \texttt{RC1} is increased above its default value the inner boundary is moved to deeper layers and thus the solutions are forced into the deep-envelope regions where the pulsation equations become stiff. For some value of \texttt{RC1} the solution using any of these methods will become unstable and start to increase in amplitude in the inwards direction exemplifying the coexistence of growing and decaying local solutions and the progressive domination of the numerical solution by the exponentially growing $e^{-k_{\max}x}$ component, where $k_{\max}$ is defined in Eq.~(\ref{eq:kMax}). For AGB envelope pulsations of radial acoustic modes there is no need to enter the asymptotically stiff regime of Eq.~(\ref{eq:basicEq}) but it serves the purpose of testing how each of the integration methods and tracking transformation behave when forced into regions with \(k_{\max}\gg 1\).

\begin{figure}[htb]
	\includegraphics{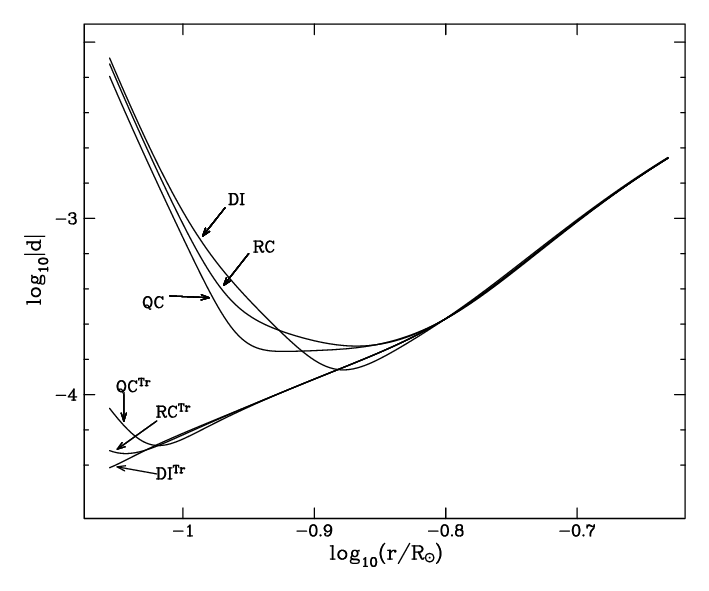}
	\FigCap{The logarithm of the displacement perturbation, normalized to unity at the surface, for different integration methods in the region of asymptotic mode decay. The value $\texttt{RC1}=50$ was chosen to extend the integration into deep envelope layers where the pulsation equations become stiff. The use of tracking helps to stabilize the solution to deeper layers.}
	\label{fig:Fig3}
\end{figure}

To illustrate the behavior of solutions to Eq.~(\ref{eq:basicEq}) when the inner boundary is moved inwards a plot of \(\log_{10}|d(x)|\) is presented in Fig.~\ref{fig:Fig3} for \(\texttt{RC1}=50\). The curves for $|d|$, for the non-tracked integrators, start to diverge far away from the imposed boundary, around the region $x=\log(r/R_{\odot})\approx -0.7$, hence negative impact of admission of divergent solutions is not limited to the close vicinity of the boundary. For $x<-0.9$ the untracked solutions start to diverge. This is also a reason why usually the boundary is placed at $\texttt{RC1}=5$ which for this mode corresponds to \(x=-0.76\). However for the tracked integrators the region of divergence of $|d|$ curves is moved deeper and it occurs around $x\approx -1$, but ultimately the tracked solution also start to diverge, provided \texttt{RC1} is set sufficiently high. The entropy perturbation is a more sensitive indicator of mode amplitude divergence, but its divergence becomes visible for the same values of \texttt{RC1} as for $|d|$. 

\begin{figure}[htb]
	\includegraphics{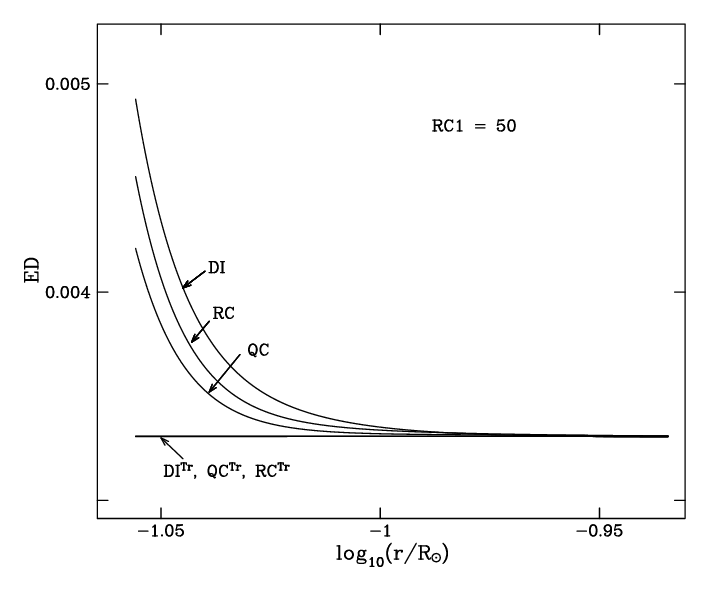}
	\FigCap{Behavior of dissipation integral $ED$ near the inner boundary for the same \texttt{RC1} as in Fig.~\ref{fig:Fig3}. It may be seen that while the $ED$ diverges close to the inner boundary for the non-tracked integrators, it remains essentially constant and identical for the three tracked ones.}
	\label{fig:Fig4}
\end{figure}

It is interesting to note that the tracking does stabilize (at least over a certain range beyond the default value of $\texttt{RC1}$) the amplitudes of perturbations allowing the expected asymptotic behavior to persist deeper into the envelope. It also improves behavior of quadratic forms, like of the dissipation integral $ED$ which depends on products of perturbations like $p^*s$. In Fig.~\ref{fig:Fig4} the behavior of $ED$ for regions of the inner envelope is shown. For the solutions using tracking the $ED$ is stable, while for the non-tracked integrators it starts to diverge near the boundary if \texttt{RC1} is set too high.

The level behavior of the $ED$ for solutions with tracking enabled implies that the phase relations between the perturbations are preserved.

It thus follows, that while tracking by itself does not prevent the solutions from becoming dominated by the divergent terms that appear when the inner boundary is placed deeper than the default level, it can be used to improve the stability of solutions in the deep-envelope regions.

\subsection{Spillover diagnostics and impending solution failure}
At the boundary the solution should reside in the 2D subspace determined by boundary conditions. However due to finite numerical precision of the integration the components of the solution may appear also in the complementary 2D subspace. In Zalewski (2026) the measure of the projection of the solution vector at the boundary onto the unwanted subspace was introduced. The amount of spillover (or the projection of the solution vector onto the unwanted subspace) may be quantified using Eq.~(9) in Zalewski (2026). Since the unwanted subspace has (in our case) two dimensions, then there are two spillover measures $\eta_{i=1,2}$. In what follows we will use the spillover norm $\eta$ defined as: $\eta=\sqrt{|\eta_1|^2+|\eta_2|^2}$. We will be interested in the measure of spillover at the inner boundary, because this is the region where the pulsation equations become stiff and the solution accuracy deteriorates. At the outer boundary the spillover measure is $\sim 10^{-15}$, so it is of no importance. The first component typically dominates $\eta$ at the inner boundary, as it represents the spillover onto the fast branch (large $k$) of the local pulsation matrix.

\begin{figure}[htb]
	\includegraphics{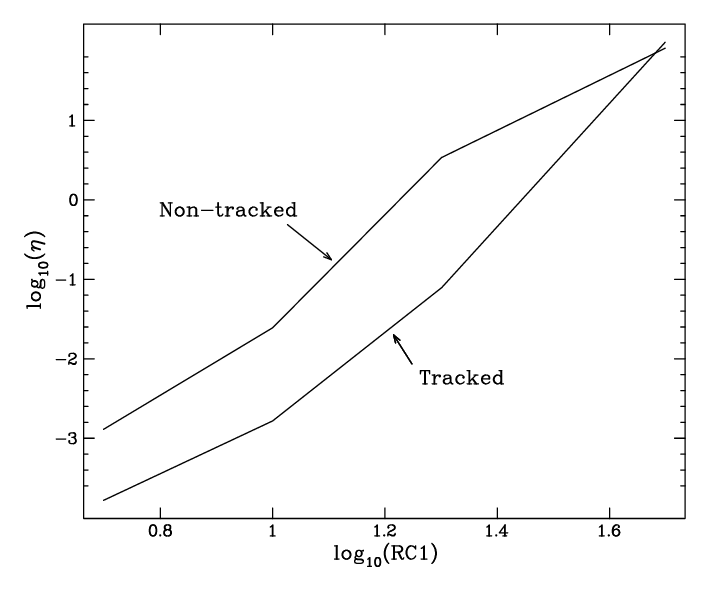}
	\FigCap{Average spillover into the unwanted inner-boundary subspace, measured by $\eta$, as a function of $\log(\texttt{RC1})$. The values are shown separately for tracked and untracked integrations and are averaged over the $\DI$, $\QC$ and $\RC$ methods. The values are for the $S^+_1$ strange mode. It may be seen that tracking reduces the spillover by about an order of magnitude for  $\texttt{RC1}<50$.}
	\label{fig:Fig5}
\end{figure}

The logarithm of spillover for the tracked and un-tracked solutions at the inner boundary is presented in Fig.~\ref{fig:Fig5}. From this figure it may be seen that the spillover is small for the default value of $\texttt{RC1}=5$. It increases with the increase of $\texttt{RC1}$. 

The spillover may be also considered as a measure of the ratio of the projection of the solution vector onto the dominant direction in the unwanted subspace at the boundary to the projection onto the dominant direction in the expected subspace. So spillover of order unity means the solution vector is distributed nearly equally between the expected subspace at the boundary and the complementary, unwanted one. Hence $\texttt{RC1}$ should be chosen so that $\eta\ll 1$. 

Integrating the solution using the tracking transformation reduces the spillover by about an order of magnitude and thus improves preservation of the intended boundary subspace. However for $\texttt{RC1}>20$ it is seen that the spillover for tracked and untracked solutions approaches comparable values and becomes unacceptably large, indicating progressive loss of solution reliability with the increase of the depth of the inner boundary.

\begin{figure}[htb]
	\includegraphics{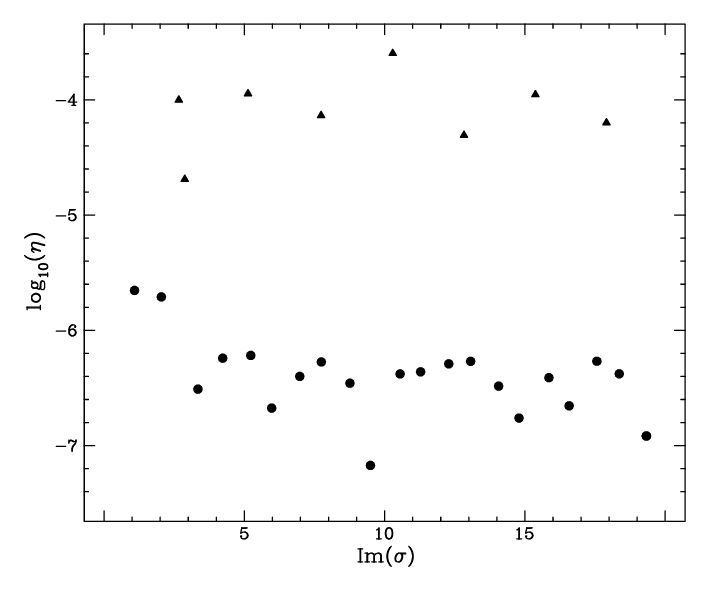}
	\FigCap{The spillover values $\eta$ for the modes seen in Fig.~\ref{fig:Fig3} for the $\DI^{\Tr}$ integrator using phase-controlled mesh. Dots mark the spillover for p-modes, while triangles for strange modes. It is seen that the spillover does not change with the increase of pulsation frequency.}
	\label{fig:Fig6}
\end{figure}

In Fig.~\ref{fig:Fig6} the values of spillovers are plotted as a function of the mode pulsation frequency for the $\DI^{\Tr}$ with $\texttt{RC1}=5$ and the same model parameters as given in section 5.1. It may be seen that the spillover values at the inner boundary are of the order of $10^{-7}$ for p-modes and they are higher, though still acceptable, for the strange modes ($10^{-4}$). While the Fig.~\ref{fig:Fig6} is for the particular integrator the results for other forms of the integrators are essentially the same. This means that for the mode integration in the considered frequency range any of the analyzed integration methods is capable of transporting the subspace defined by the outer boundary conditions to the inner boundary. It also shows that the strange modes are more difficult to integrate but still can be reliably determined.
	
This indicates that Eq.~\ref{eq:basicEq} may be solved using forward integration either by the direct approach ($\DI$) or by the nonlinear methods ($\RC$ or $\QC$). In particular, with tracking transformation enabled, the fidelity of the transported subspace is sufficient for reliable determination of the pulsation determinant through matching the subspace propagated from the surface with the subspace allowed by the inner boundary conditions.

However the comparison of spillovers for p-modes and strange modes, as well as the increase of spillover with the increase of the depth of the inner boundary indicates that in the deep-envelope the conditions lead to the deterioration of the performance of the integrator. This process can easily be quantified and controlled using \texttt{RC1} and $\eta$. 

There is, however, one additional aspect that is worth stressing. The visible mode behavior may remain apparently regular even when the transport of the intended subspace has already deteriorated. This is evidenced in Fig.~\ref{fig:Fig3}, where for  $\texttt{RC1}=50$, i.e. ten times the default value, the displacement amplitude $|d|$ still shows smooth decay for the integrators with tracking enabled and the tracked solutions preserve smooth behavior of the dissipation integral. 

This all points to the importance of the $\eta$ spillover measure as it is sensitive to the deterioration of the transported solution geometry. Progressive contamination of the propagated subspace ultimately affects the reliability of the computed eigenmodes. The $\eta$ should be used in conjunction with \texttt{RC1} and the analysis of local pulsation matrix to determine proper placement of the inner boundary.

Thus the presented approach enables the verification of the assumed value of \texttt{RC1} and provides a quantitative criterion for evaluating the reliability of the computed eigenmodes and the placement of the inner boundary.

\section{Conclusions}

In this paper we have studied the numerical reliability of nonadiabatic radial linear pulsation integrations in AGB/post-AGB envelopes using the well established methods: of direct integration of pulsation equations ($\DI$), and a Riccati method ($\RC$) introduced by Glatzel and Gautschy (1992), as well as a nonlinear integration method - $\QC$ based on continuous renormalization (Humphreys and Zumbrun, 2005). We have performed a comparison of frequency spectra and eigenmodes obtained using these methods, and refinements that may be obtained by introducing the tracking transformation. We have also analyzed the effects of different mesh types and inner boundary placement on the computation of radial modes.

We have found that for the nonlinear methods ($\QC$ and $\RC$) the spectra of frequencies are sensitive to mesh type and for lower resolution mesh types the minimum singular value maps exhibit ridge-like structures. These ridges diminish with improvement of mesh resolution and disappear for phase-controlled step-size meshes. Such meshes enable the $\RC$ and $\QC$ methods to produce consistent spectra which agree at the level $\sim 10^{-7}$.

When the mesh is sufficiently resolved the three methods, with or without tracking enabled, produce consistent spectra of pulsation frequencies. The eigenfrequencies are therefore found to be robust characteristics of the AGB/post-AGB envelope acoustic modes. However the eigenfunctions and the geometry of the propagated solution subspace are sensitive to both the use of tracking transformation and to the placement of the inner boundary.

Visually acceptable eigenfunctions and stable frequencies do not necessarily imply reliable preservation of the intended propagated solution subspace. To quantify this effect we have introduced the spillover measure $\eta$, which estimates the degree of contamination of the transported subspace by components belonging to the complementary boundary subspace.

In addition to measuring solution quality, $\eta$ provides a practical diagnostic for verifying the choice of \texttt{RC1} and thus of the placement of the inner boundary. We have established that $\texttt{RC1}=5$ is a suitable choice for the modes studied, and found that $\eta\ll 1$ for them. The tracking improves preservation of the propagated surface subspace, but does not ultimately prevent contamination growth if the boundary is be placed too deep in the envelope.

The presented approach replaces several commonly adopted numerical assumptions by explicit quantitative diagnostics of solution reliability. 
The results demonstrate that reliability of nonadiabatic pulsation calculations should be assessed not only through eigenmode convergence but also through diagnostics of transported-subspace fidelity in the asymptotic deep-envelope region. At the same time, the results show, that it is possible to obtain reliable eigenfrequencies and eigenmodes using any of the three considered integration methods.

\end{document}